\documentclass[dvips, 12pt]{article}
\usepackage{beamerarticle}

\usepackage[latin1]{inputenc}
\theoremstyle{plain}
\swapnumbers\newtheorem{stelling}{Theorem}[section]
\newtheorem{hulpstelling}[stelling]{Hulpstelling}
\newtheorem{gevolg}[stelling]{Gevolg}
\newtheorem{gevolgen}[stelling]{Gevolgen}
\newtheorem{eigenschap}[stelling]{Property}
\newtheorem{voorbeeld}[stelling]{Voorbeeld}
\newtheorem{voorbeelden}[stelling]{Voorbeelden}
\newtheorem{opmerking}[stelling]{Opmerking}
\newtheorem{opmerkingen}[stelling]{Opmerkingen}
\newtheorem{definitie}[stelling]{Definition}
\newtheorem{toepassing}[stelling]{Toepassing}

\newtheorem{definities}[stelling]{Definities}

\newtheorem{notatie}[stelling]{Notatie}

\def\bb{\begin{proof}}
\def\eb{\end{proof}}
\def\bs{\begin{stelling}}
\def\es{\end{stelling}}
\def\bhs{\begin{hulpstelling}}
\def\ehs{\end{hulpstelling}}
\def\bg{\begin{gevolg}}
\def\eg{\end{gevolg}}
\def\be{\begin{eigenschap}}
\def\ee{\end{eigenschap}}
\def\bgn{\begin{gevolgen}\hfill\begin{enumerate}}
\def\egn{\end{enumerate}\end{gevolgen}}
\def\bv{\begin{voorbeeld}\rm}
\def\ev{\end{voorbeeld}}
\def\bbv{\begin{belangrijkvoorbeeld}\rm}
\def\ebv{\end{belangrijkvoorbeeld}}
\def\bbg{\begin{belangrijkgevolg}\rm}
\def\ebg{\end{belangrijkgevolg}}
\def\bbgn{\begin{belangrijkegevolgen}\rm}
\def\ebgn{\end{belangrijkegevolgen}}
\def\bt{\begin{toepassing}}
\def\et{\end{toepassing}}
\def\bn{\begin{notatie}\rm}
\def\en{\end{notatie}}
\def\bvn{\begin{voorbeelden}\hfill\begin{enumerate}\rm}
\def\evn{\end{enumerate}\end{voorbeelden}}
\def\bo{\begin{opmerking}\rm}
\def\eo{\end{opmerking}}
\def\bon{\begin{opmerkingen}\hfill\begin{enumerate}\rm }
\def\eon{\end{enumerate}\end{opmerkingen}}
\def\bd{\begin{definitie}\rm}
\def\ed{\end{definitie}}
\def\bds{\begin{definities}\rm}
\def\eds{\end{definities}}

\def\nn{\nonumber}
 
\title[]{General Centrality in a hypergraph}
\titlegraphic{\begin{figure}
\flushleft
\includegraphics[scale=0.4]{GBI-template-cropped.png}
\end{figure}}
\author{Evo Busseniers}
\institute{GBI}

\begin{document}

\maketitle
\begin{frame}[<+->]
\end{frame}

The goal of this paper is to present a centrality measurement for the nodes of a hypergraph, by using existing literature which extends eigenvector centrality from a graph to a hypergraph, and literature which give a general centrality measurement for a graph. We'll use this measurement to say more about the number of communications in a hypergraph, to implement a learning mechanism, and to construct certain networks. 
\section{Introduction} \label{intro}

Network theory is a well established discipline. Usually the presumptions are that there is a one-to-one connection between nodes, by edges. But this is sometimes too simple for reality. We meet up with other people in groups, which has other dynamics then each of its members meeting everyone else separately. On the internet, we post messages on forums, which lead to other results then emailing someone. \\
A hypergraph formalizes this idea. It is a generalization of a graph where an edge can connect more then two edges. A hyperedge leads to different behavior then a clique (where all the nodes of the hyperedge would be connected one on one to each other). For example, a message can only be send to all or none of the members of a hyperedge, depending on whether it is passed trough the hyperedge or not.\\

The reason I am interested in a centrality measurement is to know which nodes play an important role. I want to see whether there is hierarchy in the network, whether there are nodes who have more to say or more control.\\
I liked the paper of Bonacich about general centrality \cite{bonacich_power_1987} because it gives one parameter which corresponds to different sorts of centrality. Usually, a centrality measurement looks either to the local structure ( by the degree) or to the global, the whole network (by the eigenvector centrality). The parameter in the general centrality measurement says how local or global you look. In the eigenvector centrality, the more central your neighbors are, the more central you will be. In some real life situations however, we see the opposite. In a market for example, the more central your neighbors are, the more resources they will already have, and thus the more difficult it will be for you to trade with them. Thus you will be less central. This is also taken into account in the general centrality measurement, it corresponds to the case when the parameter is negative. \\
This general centrality measurement also has a nice intuitive explanation. The centrality of a node can be seen as the number of communications starting from that node. The parameter is then equal to the chance a message is passed. Thus the bigger the parameter, the further in the network you will look.\\

I will now define a hypergraph, to be able to define all these concepts.

\frametitle{Hypergraph}

The idea of a hypergraph is to extend a graph so that edges can connect more than two nodes. It is defined as follows: 
\bd An (undirected) \textbf{hypergraph} is a couple of two sets $(V,E)$, with an element $E_j \in E$ a subset of $V$. An element $v_i \in V$ is called \textbf{node}, an element $E_j \in E$ is called \textbf{(hyper)edge}.
\ed
We can represent this by a $|V|\times|E|$-matrix $R$, where $R_{ij}=1$ if $v_i \in E_j$, and $0$ otherwise. All edges have the same strength this way, if we want the edges to have weights, we'll work with a matrix of weights $W$, with $W_{ij} \in \left[0,1 \right]$ the weight of vertex $v_i$ in edge $E_j$. In an undirected hypergraph, this is the same as the weight from $E_j$ to $v_i$. This won't be the case in a directed hypergraph, which is defined as follows:
\bd A directed \textbf{hypergraph} is a couple of two sets $(V,E)$, with an element $E_j \in E$ a couple $(I_{j}, O_{j})$ of two subsets of $V$. $I_j$ is called the set of \textbf{input nodes} and $O_j$ is called the set of \textbf{output nodes}.
\ed
A weighted directed hypergraph can be represented by two matrices: a $|V|\times|E|$-matrix $W$, giving the weights from vertices to edges, and a $|V|\times|E|$-matrix $Z$, which represents the weights from edges to nodes. In the following we will work with an undirected hypergraph, note that this can easily be extented to a directed hypergraph by using $Z$ instead of $W^T$.\\
A hypergraph can be pulled back to a standard graph. One way is to represent the hypergraph by a bipartite graph with $2$ kind of nodes: one sort is the nodes of the hypergraph, the other sort consists of the hyperedges of the hypergraph. Two nodes of different sorts are connected if the node is in the hyperedge in the hypergraph structure.\\
Another way to build a standard graph out of a hypergraph is to put an edge between all nodes which are in the same hyperedge. Consider two nodes $v_i$ and $v_k$ which are in the same hyperedge $E_j$. The contribution of $E_j$ to the weight $a_{ik}$ between $v_i$ and $v_k$ is $w_{ij} w_{kj}$. The total weight is obtained by summing over all the hyperedges which contain both $v_i$ and $v_k$, thus:
	\[ a_{ik}=\sum_{j=1}^{|E|} w_{ij} w_{kj}
\]
or in matrix notation:
	\[ A= WW^T
\]
For a directed hypergraph, we use $Z$ instead of $W^T$, thus
\[ A= WZ
\]

Note that there is some information lost: it is impossible to know in this representation which nodes are in the same hyperedge. \\

A hypergraph can have different topologies. To differentiate between them, we can extrapolate from the work already done in standard graphs. A common way to differentiate between different topologies, is to take different distributions of a property 
  of the nodes. We'll look at two properties
 : the degree and the centrality. First, we'll look at the eigenvector centrality, next, we'll introduce a more general measurement $\textbf{c}(\alpha,\beta)$, which depends on some parameters. Then we'll explain the link between the general centrality and the number of communications send by a node.\\



\section{Degree}
What is mostly done first, is to look at the degree. A hypergraph has two sorts of degree: the degree of a node, which is the number of hyperedges it is contained in, and the degree of a hyperedge, which is the number of nodes it contains. \\
\section{Centrality}

\subsection{Eigenvector centrality}		

What we will represent here, basicly comes from a paper of Volpentesta \cite{camarinha-matos_eigenvector_2010}. The idea behind the eigenvector centrality is that a node is more central as his neighboring nodes are more central. In a standard graph, the eigenvector centrality $e_i$ of a node $i$ is:
	\[\lambda e_i= \sum_j W_{ij} e_j
\]

with $W$ the matrix of the weights of the edge between two nodes ($0$ if there is no edge). $\lambda$ is just a factor so that the equations have a solution. In matrix notation this looks as follows:
	\[\lambda \textbf{e}= W \textbf{e}
\]

This is an eigenvector equation: the solutions $\textbf{e}$ are the eigenvectors of $W$, with $\lambda$ its eigenvalue. Due to the theorem of Perron-Frobenius, the eigenvector associated with the largest eigenvalue has only positive entries.\\

To extend this to a hypergraph, we will assign a centrality both on the nodes and the edges. A node is more central as the edges it is contained in are more central, and analog for the edges. Thus, the centrality $x_i$ of node $i$ is:
 	\[c_1 x_i=\sum_j W_{ij} y_j
\]
while the centrality $y_j$ of an edge $j$ is:
	\[c_2 y_j=\sum_i W_{ij} x_i .
\]
Or, in matrix notation:
	\begin{eqnarray} 
	c_1 \textbf{x}=  W \textbf{y} \nonumber \\
	c_2 \textbf{y}=  W^T \textbf{x} \nonumber
\end{eqnarray}

or, written in equations with only $\textbf{x}$ or $\textbf{y}$:
\begin{eqnarray} 			                              
 WW^T \textbf{x}= \lambda \textbf{x} \nonumber \\
 W^TW\textbf{y}=\lambda \textbf{y} \nonumber  \\
 \mbox{with } \lambda= c_1c_2      \nonumber
\end{eqnarray}

\begin{frame}[<+->]
\subsection{General centrality}		

In a graph, a general centrality measurement for a node $i$ is defined as follows \cite{bonacich_power_1987}:

			\[c_i(\alpha,\beta)= \alpha \underbrace{\sum_j W_{ij}}_{\mbox{degree}}+\beta \underbrace{\sum_j{ c_j W_{ij}}}_{\mbox{eig. centr.}}\] 
			
or in matrix notation:

\begin{eqnarray}
\textbf{c}(\alpha,\beta)=\alpha W \textbf{1}+\beta W \textbf{c}(\alpha,\beta)\nn \\
\Rightarrow (I-\beta W) \textbf{c}(\alpha,\beta)= \alpha W \textbf{1} \nn \\
\Rightarrow\textbf{c}(\alpha,\beta)= \alpha (I-\beta W)^{-1}W\textbf{1} \label{centralroot}
\end{eqnarray}

The above formula is only defined if the inverse $(I-\beta W)^{-1}$ is defined, thus if $\mbox{det}(I-\beta W)\neq 0$. We have, if $\beta \neq 0$:
\begin{eqnarray}
\mbox{det}(I-\beta W)&=& \mbox{det}(\beta (\frac{1}{\beta}I- W)) \nn \\
              &=& \mbox{det}(-\beta(W-\frac{1}{\beta}I))  \nn \\
              &=& (-\beta)^n \mbox{det}(W-\frac{1}{\beta}I) \nn
\end{eqnarray}

Since the eigenvalues of $W$ are the solutions of the equation $\mbox{det}(W-\lambda I)=0$, 
\[(-\beta)^n \mbox{det}(W-\frac{1}{\beta}I)=0\]
 if $\frac{1}{\beta}$ is an eigenvalue. This is equal to $\beta=\frac{1}{\lambda}$, with $\lambda$ an eigenvalue.\\ 
If $\beta =0$, we simpy have $I^{-1}=I$, thus the inverse is defined. This case means the centrality of a node is just $\alpha$ times his degree.
  We will use this result further on, so we put it in a property:
\be \label{inverse} $(I-\beta W)^{-1}$ is defined, if and only if $\beta \neq \frac{1}{\lambda}$, with $\lambda$ an eigenvalue.
\ee
 
If $\beta = \frac{1}{\lambda}$, we got the eigenvector centrality if $\alpha=0$.\\

The idea behind the formula is that the centrality partially depends on the local situation, measured by the degree, and partially on the global situation, measured by the eigenvector centrality. The bigger the absolute value of $\beta$, the more globally the centrality is. The sign of $\beta$ tells whether the neighbours of a node has a positive or a negative effect on that node: if $\beta$ is positive, the more central your neighbours are, the more central you will be, while if $\beta$ is negative, the more central your neighbours are, the less central you will be. An example of the second case is a market, where you're punished if you have a central neighbour, who is more difficult to trade with.\\
As can be seen in the above formula, $\alpha$ is just a scaling factor for the centrality, thus it has no effect on the distribution. Usually $\alpha$ is chosen such that 
\[\sum_i{c_i(\alpha,\beta)^2}=|V|  \]

Thus, $c_i(\alpha,\beta)=1$ means position $i$ has an average centrality.\\
\vspace{1cm}

The extension to a hypergraph is similar to the previous subsection: the centrality of a node $i$ is defined as:
		
		\[x_i=\alpha_1 \sum_j{W_{ij}}+\beta_1\sum_j W_{ij} y_j\]
		
And for an edge $j$:
		
        \[y_j=\alpha_2 \sum_i W_{ij}+ \beta_2 \sum_i W_{ij} x_i\]

In matrix notation:
\begin{eqnarray}			                                                            
 \textbf{x}= \alpha_1 W \textbf{1}+ \beta_1 W \textbf{y} \label{nodecentral} \\
 \textbf{y}=\alpha_2 W^T \textbf{1}+ \beta_2 W^T \textbf{x} \nonumber
\end{eqnarray}

Or notated with $x$ and $y$ in seperate equations:
\begin{eqnarray} 
\textbf{x}=(I-\beta_1\beta_2 WW^T)^{-1}W(\alpha_1\textbf{1}+\beta_1 \alpha_2 W^T \textbf{1}) \label{hypercentraut} \\
\textbf{y}=(I-\beta_1\beta_2 W^TW)^{-1}W^T(\alpha_2\textbf{1}+\beta_2 \alpha_1 W \textbf{1}) \nonumber
\end{eqnarray}

If $\beta_1\beta_2 \neq \frac{1}{\lambda}$, with $\lambda$ an eigenvalue of $WW^T$ or $W^TW$, the inverse is defined (\ref{inverse}).\\

If $\alpha_1=\alpha_2$, this is again a scaling factor. Since there is no straightforward explanation for the $\alpha$'s which characterize different kinds of centrality, we will often work with this assumption. There is no possibility to make the average centrality of both nodes and edges equal to $1$ with one scaling factor. We'll often take $\alpha_1=\alpha_2=1$ for simplicity.\\
			
\subsection{Communication in graph}
			
A theorem we will use in the following two chapters is:
\bs \label{sum}
$(I-cA)^{-1}=\sum_{k=0}^{+ \infty}(cA)^k$, with $A$ a symmetric matrix with positive real entries, and $c$ a constant fullfilling $|c|<\frac{1}{\lambda_{max}}$, with $\lambda_{max}$ the biggest eigenvalue of $A$.
\es
\bb
First note that it follows from \ref{inverse} that this inverse is defined, since
\[c < \frac{1}{\lambda_{max}} \leq \frac{1}{\lambda}
\]
Thus $c\neq \frac{1}{\lambda}$. \\
\bigskip 

Define 
\[ S_n=\sum_{k=0}^{n}(cA)^k. \]
Then 
\[ cA S_n=\sum_{k=0}^{n}(cA)^{k+1}=\sum_{k=1}^{n+1}(cA)^k \] 
and
\begin{eqnarray}
             S_n-cA S_n&=&I-(cA)^{n+1} \nn \\
 \Rightarrow (I-cA)S_n &=&I-(cA)^{n+1} \nn \\
 \Rightarrow       S_n &=&(I-cA)^{-1}(I-(cA)^{n+1}). \nn
\end{eqnarray} 
Thus we got:
\begin{eqnarray}
\sum_{k=0}^{+ \infty}(cA)^k&=&\lim_{n\rightarrow +\infty}{S_n} \nn \\
                           &=&\lim_{n\rightarrow +\infty}{(I-cA)^{-1}(I-(cA)^{n+1})} \nn \\
                           &=&(I-cA)^{-1} \lim_{n\rightarrow +\infty}{(I-(cA)^{n+1})} \nn \\
                           &=&(I-cA)^{-1} (I-\lim_{n\rightarrow +\infty}{(cA)^{n+1}}) \nn 
\end{eqnarray}
Thus, if we can prove that
\[\lim_{n\rightarrow +\infty}{(cA)^{n+1}}=\lim_{n\rightarrow +\infty}{(cA)^{n}}=\mbox{\begin{large}
0
\end{large}},
\]
with \begin{large}0\end{large} the zero matrix, we are there. Since $A$ is a symmetric matrix with real entries, there exists an orthogonal matrix $U$ such that $A = UDU^T$, with $D$ a diagonal matrix with the eigenvalues in the diagonal. Thus
\[ A^n=(UDU^T)^n=UDU^T UDU^T... UDU^T =U D^n U^T,
\]
where the fact that $U$ is an orthogonal matrix, meaning $U^TU=I$, is used. So we got:
\[ \lim_{n\rightarrow +\infty}{(cA)^{n}}=\lim_{n\rightarrow +\infty}{c^n U D^n U^T}.
\]
We have:
\begin{eqnarray}
c^n U \begin{pmatrix} \lambda_1^n & & \\  & \ddots &  \\ & & \lambda_n^n \end{pmatrix} U^T &\leq & c^n U \begin{pmatrix} \lambda_{max}^n & & \\  & \ddots &  \\ & & \lambda_{max}^n \end{pmatrix} U^T \nn \\
&=&c^n \lambda_{max}^n U I U^T \nn \\
&=& c^n \lambda_{max}^n I. \nn
\end{eqnarray}
Thus, if 
\[\lim_{n\rightarrow + \infty}c^n \lambda_{max}^n=0,
 \]
we are there. This is the case if
\[ -1<c \lambda_{max} < 1,
\]
thus
\[ |c| < \frac{1}{\lambda_{max}} ,
\]
which is one of the conditions and thus proves our theorem.
 \eb                                    

\end{frame}

\begin{frame}[<+->]

It follows from this theorem that there is another way to write the centrality in a graph, if $|\beta|<\frac{1}{\lambda_{max}}$, which we will assume from now on. Starting from \eqref{centralroot}, we got:

\begin{eqnarray}
\textbf{c}(\alpha,\beta)&=&\alpha (I-\beta W)^{-1}W\textbf{1}  \nn \\
                        &=^{}& \alpha (\sum_{k=0}^{+\infty}(\beta W)^k) W\textbf{1} \nn\\                        
                        &=& \alpha \sum_{k=0}^{+\infty}{\beta^k W^{k+1}\textbf{1}}  \nn
\end{eqnarray}

$\beta$ can be seen as the chance a message is passed by a node, thus $0\leq  \beta \leq 1$. In this interpretation, $\textbf{c}(1,\beta)$ can be seen as the number of communications starting from each node: we got 
\[\textbf{c}(1,\beta)= \sum_{k=0}^{+\infty}{\beta^k W^{k+1}\textbf{1}} \] 
$W\textbf{1}=$ degree of each node $=$ number of communications of length $1$;\\
$\beta W^2 \textbf{1}$= number of communications of length 2;\\
and so on. Thus the sum is the total number of communications starting from each node.\\
In the challenge propagation model\cite{heylighen_challenge_2012}, $\beta$ is the chance a challenge get selected by an agent.\\

The absolute value of $\beta$ determines the neighbourhood taken into account to calculate the centrality,$(1-\beta)^{-1}$ is the radius of this neighbourhood. After all, the expected length of a communication is
\[1+\beta+ \beta^2+...=\sum_{k=0}^{+\infty}\beta^k=\frac{1}{1-\beta}\]
The last equation holds since $\beta<1$ (the equation is a simpler case of \ref{sum}).

\end{frame}

\begin{frame}[<+->]
\subsection{Communication in hypergraph}
Using theorem \ref{sum}, we can write the centrality of the nodes in a hypergraph in a similar way as done above for a graph. This is possible if $|\beta_1 \beta_2|<\frac{1}{\lambda_{max}}$, which we will assume from now on. Starting from \eqref{hypercentraut}, we got:

\begin{eqnarray}
\textbf{x}&=& (I-\beta_1\beta_2 WW^T)^{-1}W(\alpha_1\textbf{1}+\beta_1 \alpha_2 W^T \textbf{1}) \nn \\
          &=& (\sum_{k=0}^{+ \infty}(\beta_1 \beta_2 W W^T)^k) W(\alpha_1\textbf{1}+\beta_1 \alpha_2 W^T \textbf{1})\nn \\          
          &=& \alpha_1 \sum_{k=0}^{+\infty}{(\beta_1 \beta_2 WW^T)^k W\textbf{1}}+\alpha_2\beta_1 \sum_{k=0}^{+\infty}{(\beta_1 \beta_2 )^k (WW^T)^{k+1}\textbf{1}} \nn
\end{eqnarray}

We know (Section \ref{intro} Introduction) that a hypergraph can be represented by a graph with matrix $WW^T$. If we take $\alpha_1=0; \alpha_2=1$ and $\beta_1=1$ in the above equation, we got the following result:
	\[ \textbf{x}=  \sum_{k=0}^{+\infty}{( \beta_2 )^k (WW^T)^{k+1}\textbf{1}}
\]
which is the same as the centrality of the nodes of the corresponding graph of the hypergraph, thus $\beta_2$ is the chance a challenge get selected by a node.\\

\end{frame}

\begin{frame}[<+->]

In general, if we take $\alpha_1=\alpha_2=1$, then $\beta_1$ can be interpreted as the chance an edge selects a challenge (which is in general $1$), and $\beta_2$ as the chance a node selects a challenge. Then the centrality of a node is the number of communications to a hyperedge or a node, starting from that node. 
\[\textbf{x}= \underbrace{\sum_{k=0}^{+\infty}{(\beta_1 \beta_2 WW^T)^k W\textbf{1}}}_{\mbox{communications to edges}}+\underbrace{\beta_1 \sum_{k=0}^{+\infty}{(\beta_1 \beta_2 )^k (WW^T)^{k+1}\textbf{1}}}_{\mbox{communications to nodes}}
\]
To explain why this is the number of communications, we follow a communication getting spread further and further, by looking at each term from the above equation, jumping from the first sum to the second and back. This gives us following tabular:\\

 \begin{tabular}{ccc}
      $k=0$ in 1st sum & $W\textbf{1}$                       & communications to neighbour edge \\
      $k=0$ in 2ndsum  & $\beta_1 W W^T \textbf{1}$          & communications to neighbour nodes       \\
      $k=1$ in 1st sum & $\beta_1 \beta_2 WW^T W \textbf{1}$ & communications to edges at distance $2$ \\
      $\vdots$     &   \vdots                            &   \vdots
      \end{tabular}
      
A generalization of this principle could be to let the selection of a challenge depend on the challenge (or some property/category of it) and on the agent. Thus, instead of $\beta_2 \textbf{1}$, we will work with $\bar{\beta}_2(\textbf{c})$, a vector of functions working in on an element of the challenge space. It could be the same function for all agents, the same result for all challenges, or depending on the category the challenge belongs to.

\end{frame}

\begin{frame}[<+->]

\section{Topology}

Now, we can define different topologies. This is in general done by looking at different distributions of some node property
. We will look at the degree frequency and the centrality in function of the degree. We will look at networks where this function is a power-law, normal distribution or a constant function. 


\section{Learning}

There are two ways in which the network can learn: it can adapt the weights of the already existing links, or it could create new links.\\
Weights can be adapted by delta learning: if a challenge coming from a certain hyperedge $j$ is selected by a node $i$, the weight gets adapted depending how good the challenge is relaxed:

	\[W_{ij}\leftarrow W_{ij}+r(||\textbf{c}||-||f_a(\textbf{c})||),
\]
with $f_a$ the processing function, and $r$ some constant.\\
There could also be some punishment $p$ if a challenge doesn't get selected:
\[W_{ij}\leftarrow W_{ij}-p
\]

To create new links, you can add or delete nodes to existing hyperedges by variation and selection. The fitness and mutation rate of an hyperedge $E_j$ could be:
\begin{eqnarray}
fitness(E_j)= \frac{\sum ||\textbf{c}||-||f_a(\textbf{c})||}{|E_j|} \nonumber \\
mutate(E_j)\sim \frac{1}{fitness(E_j)} \nonumber
\end{eqnarray}
Then nodes can be added depending on how much they add: $\beta_1\beta_2 W W^T W$ is the chance there is communication from canditate node to edge in $1$ step. It's also possible to look further. The chance for a communication from a node to an edge is given by the matrix $\sum_{k=0}^{+\infty}{(\beta_1 \beta_2 WW^T)^k W}$, which is equal to $(I-\beta_1\beta_2 WW^T)^{-1}W$, part of \eqref{hypercentraut}. An elaboration could be to extract it with $W$, so that nodes already in the hyperedge aren't considered.\\

To construct new hyperedges, we can treat a node as a hyperedge with only one node in, and use the same method as above to add nodes to it. If we only look to the closest nodes not yet linked with the node, it is given by the matrix $\beta_1^2\beta_2 (WW^T)^{2}$. If we consider all node-node communications, it is given by the matrix $\beta_1 \sum_{k=0}^{+\infty}{(\beta_1 \beta_2 )^k (WW^T)^{k+1}}$, which is equal to $\beta_1(I-\beta_1\beta_2 WW^T)^{-1}W W^T $, part of \eqref{hypercentraut}.  An elaboration could be to extract it with $\beta_1 W W^T$, so that nodes already connected to the other node aren't considered.\\

 Another possibility is to add an hyperedge for agents working on the same challenge, or having some kind of common goal (where the output of one agent is useful for another agent), within a certain neighbourhood.\\

Right now the weight is the same for all challenges, you could let it depend on the challenge by making it a function $W_{ij}(\textbf{c})$. Another possibility is to have different weights per component, thus a challenge could be send partially.
\end{frame}

\section{Constructing a network}

I now want to create methods to construct a network. The basic mechanism will be to extend the BA-algorithm of preferential attachment. A first extension will be to use different preferences, namely the centrality, local centrality or cluster-coefficient (for a graph) instead of the degree. I also want to use the algorithm in a hypergraph. I'm not only interesting in building a network by adding nodes, I also want to strengthen existing networks by adding edges. Here, the node we start with isn't a new node, but an existing node chosen ad random or preferential by degree.\\
Thus, we can write this in one algorithm, with two parameters to represent this different choices. The first parameter, startingnode, determines whether the first node is a new node, a node chosen ad random, or a node chosen by preference of degree. The second parameter states which variable decides the preference. Thus whether it is a higher degree, centrality or local centrality which gets more often chosen as a node or edge to connect to from the starting node. The local centrality is with respect to the starting node, it is thus the chance of communication from this node to another node. The higher this chance, the more chance to connect to the other node. This only makes sense when the starting node is an already existing node, since a new node can't reach any other node.\\

When I applied this algorithm, I noticed a problem. I often received warnings that my matrix was close to singularity. When I looked to my eigenvalues, I saw that 
the biggest eigenvalue was for some reason often around $11$. Thus $\beta$ should be smaller then $1/11= 0.0909$, which is a serious restriction I didn't want to follow. I saw that indeed there was an eigenvalue which was pretty close to $1/\beta$, the cause of the warning. \\
The solution I implemented for this problem is to give all edges a weight of $0.1$, by which the largest eigenvalue drops to $1.1$, and $\beta$ can go until $0.9$. This will always be done further on.\\
But this might be a fake solution, because lowering the weights makes it harder to traverse the network. Thus we still look pretty locally. In this case the centrality is quite similar to the degree, and it isn't that useful. We'll check whether this is indeed the case in our simulations.\\


When we construct a network by adding nodes, we will always start from a network of $3$ nodes which are all connected. In general, the weights will be $0.1$, and $\beta$ will be $0.5$.

\subsection{Results}

I first checked whether my algorithm went fine, by checking whether the plots behaved as they should when I constructed a network of $103$ nodes by preferential attachment by degree. This was the case - the degree followed a power-law.\\
Then I constructed a network by preferential attachment of centrality, and checked how it behaved differently from a network created by preferential attachment by degree. I did $1000$ iterations for both networks, thus they were $1003$ nodes. The behavior was actually quite similar, the main difference was that the maximal degree and centrality was higher in the network created by preferential attachment of degree. In both networks, the centrality is linear against the degree. This confirms that lowering the weights might be a fake solution, because we are still looking pretty locally.\\
To understand this better, we look how the centrality behaves against the degree, as $\beta$ increases. We do this in a network with weights $1$ constructed by prefential attachment of the degree, because $\beta$ isn't needed in this construction. We iterated $\beta$ from $0$ to $1$, with a step of $0.1$. For $\beta=0$, we see that indeed the centrality is linear against the degree (as expected). For $\beta=0.1$, it starts to look more as a power-law. For higher values of $\beta$, the centrality is symmetrical around zero ( there are both positive and negative values). Non-zero values of centrality tend to be with nodes with a small degree. For smaller $\beta$'s, you see a star-behavior: points tend to be on three lines, one increasing, one stable, one decreasing. There are thus still quite some nodes with a normal degree and non-zero centrality. But for higher and higher $\beta$'s, the points are squashed against the wall, until only the lowest degrees have a non-zero centrality (for $\beta=1$).\\
To try to look more globally in the network, I redid the previous construction of a network (with weights $0.1$) by preferential attachment of centrality, but with $\beta=1$. The centrality increased exponentially in comparison to the degree (not linear anymore), but for the rest the behavior was quite similar to the previous run.\\
I redid the same with $\beta=  8.5433$ (I took a strange number to avoid it being equal to a $1/\lambda$). This did give interesting results. The power-law of the degree was much less steep then before (thus it was more equally distributed). The centrality wasn't much affected by the degree, except that lower degrees tended to have a bigger absolute centrality. There were also negative centralities, thus the histogram looked like a mirrored power-law or a normal distribution (can't really tell the difference). The power-law of the cluster-coefficient against the degree was much more steep. Actually, only the nodes from the beginning had a positive cluster-coefficient, of which two had a cluster-coefficient of $1$, while their degree was $2$, thus they didn't connect to anyone new during the whole run. The other node had a cluster-coefficient of $0,05$ and a degree of $430$ (pretty average). Note that in the other runs there were also only a couple of nodes with a positive cluster-coefficient, but their degree was higher. \\

When using the cluster-coefficient as the preference, we saw that there emerged three clusters around the nodes of the starting network. Only these nodes had a positive cluster-coefficient, and thus they where the only nodes new nodes attached to.\\

For a hypergraph, the results are similar.

		

\section{Conclusion}

I defined a general centrality measurement in a hypergraph. This corresponds in certain cases to the number of communications in a hypergraph. But during testing, I saw that the restriction that $\beta$ should be smaller than $1/\lambda$ is a serious one. The chance of communication is thus pretty low, which mean we look pretty locally. Note that we can still look at the number of communications for bigger $\beta$'s, but we can only write this as an infinite sum. There is a possibility that this number will become infinite, as the number of possible paths exceeds the chance they occur. Note also that we can still look at the centrality for $\beta$'s bigger than $1/\lambda$, but that there is simply no more simple explanation as the number of communications anymore. In this case there are also negative centralities possible.\\
I will now try to understand this case as best as possible. To do this, I will basicly use the formula
\[ c_i = d_i + \beta \sum_{j \in N_i^- }{c_j} + \beta \sum_{j \in N_i^+}{c_j},
\]
 with $d_i$ being the degree of $i$, $N_i^+$ the neighbors of $i$ with positive centralities, and $N_i^-$ the neighbors with a negative centrality. For positive $\beta$'s, a node with a negative centrality means it has quite some neighbors who have a negative centrality. For negative $\beta$'s, the explanation is a bit more intuitive: a negative centrality of a node means the positive centralities of its neighbors overrule the negative ones. You could see it as nodes getting their positive centrality by 'stealing' it from their neighbors, whose centrality thus gets more negative. For $\beta$ positive, neighbors reinforce each other, but having 'bad' friends also increases your chance to be 'bad'. You could see the nodes as belonging to two 'groups': the 'positives' and the 'negatives'. Nodes belonging to different groups balance each other out, while nodes belonging to the same group reinforce each other. Note that calculating the centrality of a node with $\beta$ positive, is the same as changing the sign of the centrality of all its neighbors, and using the same $\beta$ but negative. Here, we can use the explanation for negative $\beta$'s. Thus for positive $\beta$'s, the negative nodes 'steal' centrality from you, while the positive ones give some centrality to you. Note that a positive $\beta$ isn't just the reversed state of its negative $\beta$, since we arbitrary fixed the centralitities of the neighbors, and the centrality of the given node doesn't flip, it remains the same.



\bibliographystyle{plain}
\bibliography{Centref}

\end{document}